\documentclass[preprint,aps,eqsecnum]{revtex4} 
\usepackage{bm,amsfonts}

\def\YM/{Yang\discretionary{-}{}{-}Mills}
\def\FT/{Freedman\discretionary{-}{}{-}Townsend}
\def\CM/{Chapline\discretionary{-}{}{-}Manton}
\def\CS/{Chern\discretionary{-}{}{-}Simons}
\def\YMH/{Yang\discretionary{-}{}{-}Mills\discretionary{-}{}{-}Higgs}
\def\EL/{Euler\discretionary{-}{}{-}Lagrange}

\def\eqref#1{(\ref{#1})}
\def\eqrefs#1#2{(\ref{#1}) and~(\ref{#2})}
\def\eqsref#1#2{(\ref{#1}) to~(\ref{#2})}

\def\Eqref#1{Eq.~(\ref{#1})}
\def\Eqrefs#1#2{Eqs.~(\ref{#1}) and~(\ref{#2})}
\def\Eqsref#1#2{Eqs.~(\ref{#1}) to~(\ref{#2})}

\def\Ref#1{Ref.~\cite{#1}}

\def\EQ{\begin{equation}}
\def\EQs{\begin{eqnarray}}
\def\endEQ{\end{equation}}
\def\endEQs{\end{eqnarray}}

\def\newline{\hfil\break}

\def\eqtext#1{\hbox{\ #1\ }}

\def\mstrut{\mathstrut}
\def\hp#1{\hphantom{#1}}

\def\ontop#1#2{
\setbox2=\hbox{{$#2$}} \setbox1=\hbox{{$\scriptscriptstyle #1$}} 
\dimen1=0.5\wd2 \advance\dimen1 by 0.5\wd1 \dimen2=1.4\ht2
\ifdim\wd1>\wd2 \raise\dimen2\box1 \kern-\dimen1 \hbox to\dimen1{\box2\hfill}
\else \box2\kern-\dimen1 \raise\dimen2 \hbox to\dimen1{\box1\hfill} \fi }

\def\mixedindices#1#2{{\mstrut}^{\mstrut #1}_{\mstrut #2}}
\def\downindex#1{{\mstrut}^{\mstrut}_{\mstrut #1}}
\def\upindex#1{{\mstrut}_{\mstrut}^{\mstrut #1}}
\def\downupindices#1#2{{\mstrut}_{\mstrut #1}^{\hp{#1}\mstrut #2}}
\def\updownindices#1#2{{\mstrut}^{\mstrut #1}_{\hp{#1}\mstrut #2}}

\def\tensor#1#2#3{{#1}\mixedindices{#2}{#3}}

\def\vector#1#2{{#1}\upindex{#2}}

\def\id#1#2{\delta\downupindices{#1}{#2}}

\def\A#1{\vector{A}{#1}}
\def\F#1{\vector{F}{#1}}

\def\B#1{\vector{B}{#1}}
\def\H#1{\vector{H}{#1}}

\def\K#1{\vector{K}{#1}}
\def\J#1{\vector{J}{#1}}
\def\G#1#2{\tensor{G}{#1}{#2}}
\def\tG#1#2{\tensor{\tilde G}{#1}{#2}}

\def\curvA#1{\tensor{F}{#1}{A}}
\def\curvK#1{\tensor{R}{#1}{K}}

\def\dilaton#1{\vector{\phi}{#1}}
\def\wavemap#1{\vector{\varphi}{#1}}

\def\stJ#1{\vector{\tilde J}{#1}}
\def\stA#1{\vector{\tilde A}{#1}}

\def\sX#1{\vector{\xi}{#1}}
\def\vX#1{\vector{\chi}{#1}}
\def\sXvar{\delta_\xi}
\def\vXvar{\delta_\chi}
\def\sXsub#1#2{\vector{\xi_{#1}}{#2}}

\def\sXvarsub#1{\delta_{\xi_{#1}}}
\def\vXvarsub#1{\delta_{\chi_{#1}}}

\def\Kvar{\delta_K}

\def\parity{{\mathcal P}}

\def\D#1{D\downindex{#1}}
\def\der#1{\partial\downindex{#1}}
\def\covder#1{{\bf \nabla}\downindex{#1}}

\def\conx#1#2{\Gamma\updownindices{#1}{#2}}

\def\frame#1#2{\tensor{e}{#1}{#2}}

\def\Y#1#2{Y_{A,B}\downupindices{#1}{#2}}
\def\Yinv#1#2{Y_{A,B}^{-1}\downupindices{#1}{#2}}

\def\ymA{{\bf A}}
\def\ymF{{\bf F}}
\def\ymcurvA{{\bf F}_\ymA}
\def\ftB{{\bf B}}
\def\ftK{{\bf K}}
\def\ftH{{\bf H}}

\def\ymsX{{\boldsymbol \xi}}
\def\ftvX{{\boldsymbol \chi}}
\def\ymsXvar{\delta_\ymsX}
\def\ftvXvar{\delta_\ftvX}

\def\ftJ{{\bf J}}
\def\ymG{G_\ymA}

\def\chiral{U}

\def\stymA{{\tilde {\bf A}}}
\def\stymF{{\tilde {\bf F}}}

\def\U#1{U(#1)}

\def\groupB{{\mathcal G}_\vsB}
\def\group{{\mathcal G}}

\def\vsA{{\mathcal A}}
\def\vsB{{\mathcal A'}}
\def\DvsB{{\mathcal S'}}

\def\adA{ad_\vsA}
\def\adB{ad_\vsB}
\def\adDB{ad_\DvsB}
\def\adTB{ad_\vsB^{\rm T}}

\def\BintoA{h}

\def\semi{ \setbox1=\hbox{{$\times$}} \setbox2=\hbox{{$\scriptstyle |$}}
\box1\kern -1.3\wd2 \raise.2\ht2\box2 }

\def\intprod{ {\scriptstyle\rfloor} }
\def\id{\rm id}

\def\g#1#2{g\downupindices{#1}{#2}}
\def\hmap#1#2{h\downupindices{#1}{#2}}

\def\atens#1#2{a\upindex{#1}\downindex{#2}}
\def\btens#1#2{b\upindex{#1}\downindex{#2}}
\def\ctens#1#2{c\downindex{#1}\upindex{#2}}
\def\etens#1#2{e\upindex{#1}\downindex{#2}}

\def\Tctens#1#2{c\upindex{#1}\downindex{#2}}
\def\Tbtens#1#2#3{b\downindex{#1}\upindex{#2}\downindex{#3}}
\def\Tetens#1#2{e\downindex{#1}\upindex{#2}}

\def\evec#1#2{e'\mixedindices{#1}{#2}}

\def\nthL#1{\ontop{(#1)}{L}}

\def\nthsXvar#1{\ontop{(#1)}{\delta_\xi}}
\def\nthvXvar#1{\ontop{(#1)}{\delta_\chi}}

\def\frac#1#2{{\textstyle {#1 \over #2}}}

\def\Rnum{{\mathbb R}}

\def\inv{{}^{-1}}
\def\ad{{}^{\rm T}}
\def\intprod{ {\scriptstyle\rfloor} }

\def\openone{{\bf 1}}

\def\ie/{i.e.}
\def\eg/{e.g.}

\newtheorem{theorem}{Theorem}

\newtheorem{proposition}[theorem]{Proposition}

\begin{document}

% date of version: October 22, 2002

\title{ Exotic Yang-Mills dilaton gauge theories }

\author{Stephen C. Anco}
\affiliation{
Department of Mathematics, 
Brock University,
St Catharines, ON Canada L2S 3A1 }
\email{sanco@brocku.ca}

\begin{abstract}
An exotic class of nonlinear $p$-form nonabelian gauge theories is studied,
arising from the most general allowed covariant deformation of
linear abelian gauge theory for a set of 
massless $1$-form fields and $2$-form fields in four dimensions. 
These theories combine a Chapline-Manton type coupling of 
the $1$-forms and $2$-forms, 
along with a Yang-Mills coupling of the $1$-forms,
a Freedman-Townsend coupling of the $2$-forms,
and an extended Freedman-Townsend type coupling 
between the $1$-forms and $2$-forms. 
It is shown that the resulting theories have a geometrically interesting
dual formulation that is equivalent to an exotic Yang-Mills dilaton theory
involving a nonlinear sigma field. 
In particular, the nonlinear sigma field couples 
to the \YM/ $1$-form field through a generalized Chern class $4$-form term. 
\end{abstract}


\maketitle
\newpage

\section{ Introduction }
\label{intro}

There has been much recent interest in the systematic construction of
nonlinear $p$-form gauge theories
with couplings among $p$-form fields for $p\ge 2$ in $n$ dimensions. 
The construction begins with deforming 
the linear abelian gauge theory of $p$-form fields 
by addition of cubic terms in the free field Lagrangian 
and linear terms in the abelian gauge symmetries
such that gauge invariance of the theory is maintained to first-order. 
All allowed first-order deformation terms can be obtained 
as solutions of determining equations 
derived from the condition of gauge invariance. 
The full nonlinear theory consists of completing the deformation 
of the gauge symmetries and Lagrangian to all orders \cite{AMSpaper}.

A classification of first-order deformations in $n$ dimensions 
has been derived in \Ref{HenneauxKnaepen1,HenneauxKnaepen2}
by formulating and solving the relevant determining equations
in the setting of BRST cohomology 
\cite{Henneaux1,BarnichBrandtHenneaux1}. 
In particular, for $p$-form fields with $p\ge 2$, 
the allowed cubic terms for deforming the $n$-form Lagrangian 
are classified in terms of the abelian $p$-form field strengths as follows:
\FT/ \cite{FTth} type couplings (if $n\ge 3$), 
which are quadratic in the dual field strengths;
\CM/ \cite{CMth} type couplings (if $n\ge 4$), 
which are linear in the dual field strengths
and at least linear in the field strengths;
higher derivative generalizations (if $n\ge 5$), 
which are quadratic in the dual field strengths
and at least linear in the field strengths. 
When $1$-form fields are considered,
there is also the \YM/ coupling (if $n\ge 2$), 
which is linear in the dual field strength of the $1$-form. 
The full nonlinear gauge theories 
for each of these separate types of couplings
are well-known. 
However, there has been less effort to-date in the study of 
theories that combine the different types of couplings together. 
Such theories are expected to have somewhat exotic features,
as exhibited in the cases of combined \YM//\FT/ couplings 
in $n=3$ and $n=4$ dimensions 
studied in \Ref{YMpapers,YMFTpaper}. 

In this Letter, 
full nonlinear gauge theories for the cases of 
a \CM/ coupling combined with \FT/ and \YM/ couplings in $n=4$ dimensions
are explored. 
The main result is to show that 
these theories have an interesting geometrical formulation
and duality which, in general, is equivalent to 
an exotic \YM/ dilaton theory
involving a nonlinear sigma field. 
In particular, the nonlinear sigma field is found to couple 
to the \YM/ $1$-form through a generalized Chern class term
\cite{Chernclass}.

\section{ Preliminaries }

We start from the results of the analysis in \Ref{YMFTpaper}
for geometrical first-order deformations \cite{geometrical}
of linear abelian gauge theory of
a set of massless $1$-forms $\A{a}$ ($a=1,\ldots,k$)
and $2$-forms $\B{a'}$ ($a'=1,\ldots,k'$)
with field strength $2$-forms $\F{a}=d\A{a}$ and $3$-forms $\H{a'}=d\B{a'}$
on a four-dimensional manifold $M$.
(Throughout, 
$d$ is the exterior derivative operator, 
$*$ denotes the Hodge dual operator with $*^2=\pm \openone$, 
and products of $p$-forms are understood to be wedge products.
All constructions will be local, \ie/ in a single coordinate chart, 
with respect to $M$.)
The undeformed Lagrangian and gauge symmetries are given by 
\EQ
\nthL{2} = *\F{a} \F{b} \g{ab}{} + *\H{a'} \H{b'} \g{a'b'}{}
\label{linthL}
\endEQ
with coefficients $\g{ab}{},\g{a'b'}{}=diag(\pm1,\ldots,\pm1)$, 
and 
\EQ
\nthsXvar{0}\A{a} =d\sX{a} ,\quad
\nthvXvar{0}\B{a'} = d\vX{a'} ,\quad
\nthvXvar{0}\A{a} = \nthsXvar{0}\B{a'} =0 , 
\label{linthX}
\endEQ
where $\sX{a}$ is a set of arbitrary $0$-forms (\ie/ functions)
and $\vX{a'}$ is a set of arbitrary $1$-forms on $M$. 

\begin{proposition}
The separate \YM/, (extended) \FT/, \CM/ first-order deformations of
linear abelian $p$-form gauge theory ($p=1,2$) on $M$
consist of the terms \cite{HenneauxKnaepen1,YMFTpaper} 
\EQs
\nthL{3} = &&
( \frac{1}{2} \atens{}{abc} {*\F{a}} \A{b}\A{c} )_{\rm YM} 
+( -\frac{1}{2}\ctens{a'b'c'}{} {*\H{a'}} {*\H{b'}} \B{c'} )_{\rm FT} 
\nonumber\\&&
+( \btens{}{ab'c} {*\F{a}} {*\H{b'}} \A{c} )_{\rm exFT} 
+( -\etens{}{a'bc} {*\H{a'}} \F{b} \A{c} )_{\rm CM} 
\label{deformations}
\endEQs
and 
\EQs
&& 
\nthsXvar{1}\A{a} =
( \atens{a}{bc} \A{b}\sX{c} )_{\rm YM} 
+ ( \btens{a}{b'c} {*\H{b'}} \sX{c} )_{\rm exFT} ,
\\
&& 
\nthsXvar{1}\B{a'} =
( -\Tbtens{b}{a'}{c} {*\F{b}} \sX{c} )_{\rm exFT} 
+( \etens{a'}{bc} \F{b} \sX{c} )_{\rm CM} , 
\\
&&
\nthvXvar{1}\A{a} =0 ,\quad
\nthvXvar{1}\B{a'} = 
( \Tctens{a'}{b'c'} {*\H{b'}} \vX{c'} )_{\rm FT} , 
\endEQs
where the coupling constants satisfy the algebraic relations
\EQs
&&
\atens{}{a(bc)} =0 ,\quad
\atens{}{ab[c} \atens{b}{de]} =0 ,\quad
\ctens{(a'b')c'}{} =0 ,\quad
\ctens{[d'e'}{c'} \ctens{b']c'a'}{} =0 ,
\label{aceq}\\
&&
2\btens{}{a[b'|c} \btens{c}{|d']e} = \btens{}{ac'e} \ctens{b'd'}{c'} ,\quad
\etens{}{a'[bc]}=0 .
\label{beeq}
\endEQs
\end{proposition}

Additional algebraic relations on the coupling constants arise
as integrability conditions 
for the existence of second-order deformation terms
when one considers deformations that combine 
pure \YM/, (extended) \FT/, and \CM/ couplings. 

\begin{theorem}
The necessary and sufficient algebraic obstructions
for existence of mixed \YM/, (extended) \FT/, \CM/ 
second-order deformations 
are given by \cite{YMFTpaper}
\EQs
&&
2\atens{}{ab[c} \btens{b}{|d'|e]} = \btens{}{ad'b} \atens{b}{ec} ,
\label{abeq}\\
&&
\etens{}{a'c(b} \atens{c}{d)e} =0 ,
\label{eaeq}\\
&&
\ctens{d'e'}{c'} \etens{}{c'ab}
=2 \etens{}{[d'|ca} \btens{c}{|e']b} +2 \etens{}{[d'|cb} \btens{c}{|e']a} .
\label{eceq}
\endEQs
\end{theorem}

The relations \eqsref{aceq}{eceq}
impose a certain natural algebraic structure 
on the internal vector spaces, $\vsA=\Rnum^k,\vsB=\Rnum^{k'}$,
associated with the set of fields $\A{a},\B{a'}$.
Specifically, this structure is given by:
\vskip0pt
(i) 
$\atens{a}{bc}$ defines structure constants of 
a Lie bracket $[\cdot,\cdot]_\vsA$ from $\vsA\times\vsA$ into $\vsA$
such that $\g{ab}{}$ is an invariant metric on $\vsA$, \ie/
\EQs
&&
[u,v]_\vsA = - [v,u]_\vsA ,\quad
[u,[v,w]_\vsA]_\vsA + \eqtext{cyclic\ terms} =0 ,
\\&&
g(u,[v,w]_\vsA) = g([u,v]_\vsA,w) ;
\label{galgeq}
\endEQs
\vskip0pt
(ii) 
$\ctens{a'b'}{c'}$ defines structure constants of 
a Lie bracket $[\cdot,\cdot]_\vsB$ from $\vsB\times\vsB$ into $\vsB$, \ie/
\EQ
[u',v']_\vsB = - [v',u']_\vsB ,\quad
[u',[v',w']_\vsB]_\vsB + \eqtext{cyclic\ terms} =0 ,
\endEQ
while $\g{a'b'}{}$ need not be an invariant metric;
\vskip0pt
(iii)
$\btens{a}{b'c}$ defines a linear map $b(\cdot)$ 
from $\vsB\times\vsA$ into $\vsA$ 
given by, jointly, 
a representation of the Lie algebra $\vsB$ on the vector space $\vsA$
and a derivation of the Lie algebra $\vsA$, \ie/
\EQ
[b(u'),b(v')] =b([u',v']_\vsB) ,\quad
b(w')[u,v]_\vsA =[b(w')u,v]_\vsA + [u,b(w')v]_\vsA ;
\label{balgeq}
\endEQ
\vskip0pt
(iv)
$\etens{a'}{bc}$ defines a symmetric product $e(\cdot,\cdot)$ 
from $\vsA\times\vsA$ into $\vsB$
that is invariant with respect to the Lie bracket on $\vsA$
\EQ
e(u,v) =e(v,u) ,\quad
e(u,[v,w]_\vsA) =  e([u,v]_\vsA,w) 
\label{ealgeq}
\endEQ
and such that the related symmetric linear map $e^S(\cdot)$ 
from $\vsB\times\vsA$ into $\vsA$ defined by $\Tetens{a'b}{c}$
intertwines with the representation of $\vsB$ via
\EQ
e^S([u',v']_\vsB) - [e^S(u'),b^A(v')] -[b^A(u'),e^S(v')]
= \{e^S(u'),b^S(v')\} - \{b^S(u'),e^S(v')\}
\label{ealgeq'}
\endEQ
where $b^S(\cdot)= \frac{1}{2}( b(\cdot) + b\ad(\cdot) )$
and $b^A(\cdot)= \frac{1}{2}( b(\cdot) - b\ad(\cdot) )$
are the symmetric and skew parts of the representation $b(\cdot)$.

By a generalization of the algebraic analysis of \Ref{YMFTpaper}
applied to (i)-(iv), 
one obtains the following result. 

\begin{proposition}
Suppose the metrics given by $\g{ab}{},\g{a'b'}{}$ on $\vsA,\vsB$ 
are positive definite. 
Then the Lie algebra $\vsA$ is compact semisimple or abelian. 
In the semisimple case for $\vsA$, 
\Eqref{galgeq} is satisfied by the Cartan-Killing metric
$g(u,v) =-tr(\adA(u)\adA(v)) \equiv tr_\vsA(uv)$ 
where $\adA(\cdot)$ is the adjoint representation of $\vsA$,
\ie/
\EQ
\g{ab}{} = -\atens{d}{ca}\atens{c}{db} ,
\endEQ
and \Eqref{balgeq} is satisfied in terms of $\adA(\cdot)$ by 
$b(w') =\adA(\BintoA(w'))=-b\ad(w')$
with $\BintoA(\cdot)$ being a Lie-algebra homomorphism of $\vsB$ into $\vsA$,
$[\BintoA(u'),\BintoA(v')]_\vsA = \BintoA([u',v']_\vsA)$, 
\ie/
\EQ
\btens{a}{b'c} = \atens{a}{bc} \hmap{b'}{b} ,\quad
\hmap{b'}{b} \hmap{c'}{c} \atens{a}{bc} = \hmap{a'}{a} \ctens{b'c'}{a'} .
\endEQ
Moreover, \Eqrefs{ealgeq}{ealgeq'} are then satisfied by 
$e(u,v)=e'\otimes g(u,v)$
for any vector $e'$ in $\vsB$ 
orthogonal to the commutator ideal $D\vsB=[\vsB,\vsB]$,
\ie/
\EQ
\etens{a'}{bc} =\evec{a'}{} \g{bc}{} ,\quad
\evec{}{a'} \ctens{b'c'}{a'} =0
\endEQ
(and hence $e'=0$ whenever $\vsB=D\vsB$). 
However, in the abelian case for $\vsA$,
\Eqrefs{ealgeq}{ealgeq'} are instead satisfied by 
$e^S(w') = e b^S(w')$ 
for any constant $e$ if $b^S(\cdot)\neq 0$ 
or by 
$e^S(w') = g(e',w')\id_\vsA$ 
for any vector $e'$ as above if $b^S(\cdot) = 0$,
\ie/
\EQ
\Tetens{a'b}{c} = \cases{ 
e \frac{1}{2}(\btens{c}{a'b} + \Tbtens{ba'}{c}{}) 
& if $\btens{}{(c|a'|b)} \neq 0$ \cr
\evec{}{a'} \g{b}{c} 
& if $\btens{}{(c|a'|b)} = 0$ }
\label{esol}
\endEQ
which is seen from a comparison of \Eqref{eceq} with 
the symmetric part of \Eqref{beeq}. 
\end{proposition}

The aim now is to write down a class of nonlinear gauge theories
giving a complete deformation to all orders 
(in particular, there are no further algebraic obstructions).
To illustrate the nature and essential pattern for this class of theories,
we first examine the \CM/ coupling of 
an abelian $1$-form $A$ and abelian $2$-form $B$,
with field strengths $F=dA$, $H=dB$.

The Lagrangian is given by the $4$-form \cite{HenneauxKnaepen1,YMFTpaper}
$L= *F F +*(H-e FA) (H-e FA)$
where $e$ is a coupling constant. 
This Lagrangian is invariant 
under the separate abelian gauge symmetries
$\sXvar A=d\sX{}$, $\sXvar B=e F\sX{}$,
for arbitrary $0$-forms $\sX{}$,
and $\vXvar A =0$, $\vXvar B=d\vX{}$,
for arbitrary $1$-forms $\vX{}$. 
The field equations obtained from $L$ are given by 
\EQ
0=d {*F} + e ( 2F ({*H} -e {*(FA)}) -A d( {*H} -e {*(FA)} ) ) ,\quad
0=d(*H -e {*(FA)}) .
\endEQ
This theory describes a nonlinear deformation of 
abelian Maxwell/\FT/ gauge theory for $A,B$. 
It has a dual formulation in terms of a scalar field introduced 
through the $B$ field equation, 
\EQ
*(H -e FA) = d\dilaton{} .
\label{Heq}
\endEQ
Then the $A$ field equation becomes
\EQ
d {*F} = -2 e Fd\dilaton{} .
\endEQ
The $\dilaton{}$ field equation is obtained by elimination of $H$
in equation \eqref{Heq}, 
which yields
\EQ
d{*d}\dilaton{} = \mp e F F . 
\endEQ
These field equations for $\dilaton{}$ and $A$ arise equivalently
from the Lagrangian $4$-form
\EQ
L^{\rm dual} = {*F} F + (2e G \pm {*d}\dilaton{}) d\dilaton{}
\endEQ
where $G=AF=AdA$ is the abelian \CS/ $3$-form \cite{Chernref},
satisfying $dG= FF$. 
Thus, $\dilaton{}$ couples to $A$ through a \CS/ term
$G d\dilaton{} = \dilaton{} FF$ (to within a trivial exact term)
where $FF$ is the abelian Chern class $4$-form \cite{Chernref} on $M$. 

The Lagrangian $L^{\rm dual}$ has a more direct derivation 
from the nonlinear gauge theory for $A$ and $B$
by passing to a 1st order formalism
$L^{1st} = {*F}F +(2eG \pm *K)K +2B dK$
with the $1$-form $K$ being an auxiliary field variable. 
Elimination of $K$ via its field equation
$K+e{*G} = *dB$
leads to the original Lagrangian $L$ in terms of $A,B$. 
Alternatively, through the $B$ field equation 
$dK=0$, 
the introduction of $\dilaton{}$ via 
$K=d\dilaton{}$ yields the dual Lagrangian $L^{\rm dual}$
for $\dilaton{},A$. 

This 1st order formulation will now be exploited
for generalizing the \CM/ coupling to 
a set of $1$-forms $\A{a}$ ($a=1,\ldots,k$)
and $2$-forms $\B{a'}$ ($a'=1,\ldots,k'$)
with a \YM/ coupling on $\A{a}$, 
a \FT/ coupling on $\B{a'}$,
and an extended \FT/ coupling between $\A{a}$ and $\B{a'}$.

\section{ Construction of the class of nonlinear theories }

We being by writing down the nonabelian \YM/ generalization of
the 1st order Lagrangian for the extended \FT/ gauge theory of 
$1$-forms $\A{a}$ and $2$-forms $\B{a'}$,
similarly to the abelian theory discussed in \Ref{HenneauxKnaepen1}. 
The 1st order formulation uses auxiliary $1$-forms $\K{a'}$
whose role is a \FT/ nonlinear field strength. 
Let the $2$-forms $\curvA{a} = d\A{a} +\frac{1}{2}\atens{a}{bc} \A{b}\A{c}$
denote the nonabelian \YM/ field strength. 
Then the Lagrangian is given by the $4$-form
\EQ
L^{1st}_{\rm YM/exFT} =
*\J{a} \J{b} \g{ab}{} + (2\B{a'} \curvK{b'} \pm *\K{a'} \K{b'})\g{a'b'}{} ,
\endEQ
where 
\EQ
\curvK{a'} = d\K{a'} +\frac{1}{2} \ctens{b'c'}{a'} \K{b'}\K{c'}
\endEQ
and 
\EQ
\J{a} = \curvA{a} + \btens{a}{b'c} \K{b'}\A{c}
\equiv \D{K}\A{a} + \frac{1}{2}\atens{a}{bc} \A{b}\A{c} .
\endEQ
Geometrically, 
$\curvK{a'}$ is the curvature of the \FT/ connection $\K{a'}$,
while $\J{a}$ is a generalized curvature of the \YM/ connection $\A{a}$
involving a \FT/ coupling through the covariant derivative operator $\D{K}$
associated with the connection $\K{a'}$. 
(The underlying gauge groups for these connections
are the unique local Lie groups whose Lie algebras are $\vsB,\vsA$.)
This 1st order Lagrangian is invariant under 
combined \YM/ and extended \FT/ gauge symmetries. 
A \CM/ coupling is now introduced in this theory as follows. 
First, let 
$\G{a'}{A} 
= \etens{a'}{bc}( \A{b} d\A{c} +\frac{1}{3} \atens{c}{de} \A{b}\A{d}\A{e} )$
denote a nonabelian \CS/ $3$-form, 
satisfying
$d\G{a'}{A} =\etens{a'}{bc} \curvA{b} \curvA{c}$
due to property \eqref{ealgeq}, 
and next define the related $3$-form 
\EQ
\G{a'}{} 
= \G{a'}{A} -\etens{a'}{bc} \btens{c}{b'd} \A{b}\A{d} \K{b'} 
= \etens{a'}{bc} \A{b}( \D{K}\A{c}  +\frac{1}{3} \atens{c}{de} \A{d}\A{e} ) .
\endEQ
Then add the coupling term 
\EQ
L^{1st}_{\rm CM} = 2\G{a'}{} \K{b'} \g{a'b'}{}
\endEQ
to the 1st order Lagrangian,
yielding the complete Lagrangian 
$L^{1st} = L^{1st}_{\rm YM/exFT} + L^{1st}_{\rm CM}$
for the theory. 
The gauge symmetries under which this Lagrangian is invariant
(to within an exact $4$-form) are given by
\EQs
&&
\vXvar\B{a'} =
d\vX{a'} +\Tctens{a'}{b'c'} \K{b'}\vX{c'} \equiv \D{K}\vX{a'} ,
\label{vX}\\
&&
\sXvar\B{a'} = 
\etens{a'}{bc} \A{b} \D{K}\sX{c} 
-\Tbtens{b}{a'}{c}( {*\J{b}} +2\Tetens{b'd}{b} \A{d}\K{b'} )\sX{c} ,
\\
&&
\vXvar\A{a} = 0 ,\quad
\sXvar\A{a} =
\D{K}\sX{a} + \atens{a}{bc} \A{b} \sX{c} \equiv \D{K+A}\sX{a} ,
\label{sX}\\
&&
\vXvar\K{a'} = \sXvar\K{a'} = 0 .
\endEQs
Here $\vXvar$ is a \FT/ gauge symmetry,
and $\sXvar$ is a \YM/ generalization of an extended \FT/ gauge symmetry
combined with a \CM/ gauge symmetry. 
(The commutators are given by 
$[\sXvarsub{1},\sXvarsub{2}]= \sXvarsub{3}$
with $\sXsub{3}{a}=\atens{a}{bc} \sXsub{1}{b}\sXsub{2}{c}$,
and $[\sXvarsub{1},\vXvarsub{1}]=[\vXvarsub{1},\vXvarsub{2}]=0$,
to within trivial symmetries proportional to the field equations.
Hence the gauge symmetry algebra is $\vsA \times \U{1}^{k'}$.)

By elimination of $\K{a'}$ via its field equation
\EQ
\K{a'} + *\G{a'}{} - \Tbtens{b}{a'}{c} *(
\A{c} (*\J{b} + \Tetens{d'e}{b} \A{e} \K{d'}) ) 
= *\D{K}\B{a'} , 
\label{Keq}
\endEQ
we obtain a nonlinear gauge theory for $\A{a},\B{a'}$,
with the Lagrangian 
\EQ
L_{\rm YM/exFT/CM} = 
\K{a'}( \H{b'} -\G{b'}{A} )\g{a'b'}{} 
+ *\J{a} \curvA{b} \g{ab}{}
\label{L}
\endEQ
which is invariant under the previous gauge symmetries on $\A{a},\B{a'}$.
As one sees by \Eqref{Keq}, 
the 1-form 
$\K{a'}=\Yinv{b'}{a'}\intprod(*\H{b'}-*\G{b'}{A})$
in this theory
is a nonpolynomial field strength for $\B{a'}$
given in terms of the inverse of a symmetric linear map
defined on $\vsB$-valued 1-forms by 
\EQ
\Y{b'}{a'} = 
\g{b'}{a'}\cdot\,
-\Tctens{a'}{b'c'} {*(\B{c'}\cdot)}
+\Tbtens{d}{a'}{b}\btens{d}{b'c}
{*(\A{b} {*(\A{c} \cdot)})}
+( \etens{a'}{bd}\btens{d}{b'c}-\Tetens{b'b}{d}\Tbtens{d}{a'}{c} )
{*(\A{b}\A{c} \cdot)} . 
\label{Ymap}
\endEQ

\begin{theorem}
The class of nonlinear gauge theories \eqsref{vX}{Ymap}
comprise the most general deformation (to all orders)
of linear abelian gauge theory \eqrefs{linthL}{linthX}
determined by combining \YM/, extended \FT/, and \CM/ 
first-order deformations \eqref{deformations}
for a set of $1$-forms $\A{a}$ and $2$-forms $\B{a'}$ on $M$. 
\end{theorem}

This class of theories is equivalent to 
an exotic \YM/ nonabelian dilaton theory 
obtained from the 1st order Lagrangian 
by eliminating $\B{a'}$ via its field equation $\curvK{a'}=0$,
as follows. 
Since $\curvK{a'}$ is the curvature of $\K{a'}$,
we see that $\K{a'}$ is a flat connection
and hence is of form 
\EQ
\K{a'} = \frame{a'}{\mu}(\wavemap{}) d\wavemap{\mu}
\endEQ
in terms of a nonlinear sigma field $\wavemap{\mu}$ given by 
a map from $M$ into the local Lie group $\groupB$ 
determined by the Lie algebra $\vsB$, 
where $\frame{a'}{\mu}(\wavemap{})$ is a left-invariant basis of $1$-forms
(\ie/ a frame) on $\groupB$, 
with 
$2\der{[\nu} \frame{a'}{\mu]} = 
\ctens{b'c'}{a'} \frame{b'}{\mu} \frame{c'}{\nu}$. 
(This is an extension of the well-known geometrical equivalence \cite{FTth} 
of nonlinear sigma theory for $\wavemap{\mu}$ 
based on a Lie group target $\groupB$ 
and pure \FT/ gauge theory for $\B{a'}$ 
involving the Lie algebra $\vsB$ of $\groupB$.)
Consequently, once $\B{a'}$ is eliminated in the 1st order Lagrangian,
we obtain the dual Lagrangian  
\EQ
L^{\rm dual} =
*\J{a} \J{b} \g{ab}{} 
+( 2\G{a'}{} \frame{\mu}{a'}(\wavemap{}) \pm {*d}\wavemap{\mu} ) 
d\wavemap{\nu} \g{\mu\nu}{}(\wavemap{})
\endEQ
where 
\EQ
\g{\mu\nu}{}(\wavemap{}) =
\frame{a'}{\mu}(\wavemap{}) \frame{b'}{\nu}(\wavemap{}) \g{a'b'}{}
\endEQ
is the left-invariant metric on $\groupB$ 
associated with the Lie algebra metric $\g{a'b'}{}$,
and here $\J{a}$ and $\G{a'}{}$ are now defined in terms of 
the covariant derivative
$\D{K} = d +\btens{a}{b'c} \frame{b'}{\mu} d\wavemap{\mu}$.
Note $\frame{\mu}{a'}$ is the coframe of $\frame{a'}{\mu}$ on $\groupB$. 
The field equations for $\A{a}$ and $\wavemap{\mu}$,
after some simplifications using \Eqsref{galgeq}{ealgeq}, 
are given by 
\EQ
\D{A} {*\J{a}} = \frame{b'}{\mu}(\wavemap{}) d\wavemap{\mu}( 
\Tbtens{cb'}{a}{} {*\J{c}} -2\Tetens{b'c}{a} \J{c} ) , 
\endEQ
\EQ
\pm( d{*d}\wavemap{\mu} 
+\conx{\mu}{\nu\sigma}(\wavemap{}) d\wavemap{\nu} {*d}\wavemap{\sigma} )
= \frame{\mu}{a'}(\wavemap{})( 
\Tbtens{b}{a'}{c} {*\J{b}}\J{c} -\etens{a'}{bc} \J{b}\J{c} ) ,
\endEQ
where 
$\conx{\mu}{\nu\sigma}(\wavemap{}) = 
\g{}{\mu\tau}(\wavemap{})( \der{(\nu} \g{\sigma)\tau}{}(\wavemap{})
-\frac{1}{2} \der{\tau} \g{\nu\sigma}{}(\wavemap{}) )$
is the Christoffel symbol of the Lie-group metric 
$\g{\mu\nu}{}(\wavemap{})$. 
(Here, geometrically, 
$d+\conx{\mu}{\nu\sigma}(\wavemap{})d\wavemap{\nu} =\covder{g}$
is the pullback to $M$ of 
the unique torsion-free derivative operator on $\groupB$
determined by the metric.)
Thus, $\wavemap{\mu}$ couples to $\A{a}$ through 
the $4$-form terms 
$\etens{a'}{bc} \J{b}\J{c}$ and $\Tbtens{b}{a'}{c} {*\J{b}}\J{c}$.

From the manifest similarity between 
$\Tbtens{b}{a'}{c} {*\J{b}}\J{c}$
and the generalized \YM/ term $*\J{b}\J{c}\g{bc}{} \equiv L_{\rm YM}$
in the Lagrangian $L^{\rm dual}$,
the $4$-form $\Tbtens{b}{a'}{c} {*\J{b}}\J{c}$
is seen to describe a dilaton type coupling 
between $\wavemap{\mu}$ and $\A{a}$ in the field equations. 
Further discussion of the nature of this coupling is given 
at the end of the next section. 

In comparison, the $4$-form 
$\etens{a'}{bc} \J{b}\J{c}$
in the field equations
describes an exotic type of dilaton coupling, 
related to a generalized Chern class term as follows. 
Let 
\EQ
\G{a'}{K} 
= \G{a'}{A} 
- (\etens{a'}{bc} \btens{c}{b'd} + \Tbtens{c}{a'}{b} \Tetens{b'd}{c})
\A{b}\A{d} \K{b'} 
\endEQ
as given by the variational derivative of the \CM/ term 
$L^{1st}_{\rm CM}$ with respect to $\K{a'}$, \ie/ 
$\Kvar L^{1st}_{\rm CM} = 2 \G{a'}{K} \delta\K{b'} \g{a'b'}{}$. 
Likewise, let 
\EQ
\tG{a'}{K} 
= \Tbtens{b}{a'}{c} \A{c} {*\J{b}} 
= \Tbtens{b}{a'}{c} \A{c} 
{*( \D{K}\A{b} +\frac{1}{2}\atens{b}{de} \A{d}\A{e} )}
\endEQ
obtained from the \YM/ term $L_{\rm YM}$. 

\begin{proposition}
The $3$-forms $\G{a'}{K}$ and $\tG{a'}{K}$ are, jointly, 
potentials for the gauge invariant $4$-forms 
$\etens{a'}{bc} \J{b}\J{c}$ and $\Tbtens{b}{a'}{c} {*\J{b}}\J{c}$,
satisfying the relation
\EQ
\D{K}( \G{a'}{K} - \tG{a'}{K} ) = 
\etens{a'}{bc} \J{b}\J{c} - \Tbtens{b}{a'}{c} {*\J{b}}\J{c}
\label{Chernclass}
\endEQ
on all solutions $\A{a}$ of the field equations. 
This is a generalization of the relation \cite{Chernref} between 
the \CS/ $3$-form $\G{a'}{A}$
and the \YM/ Chern class $4$-form 
$\etens{a'}{bc} \curvA{b} \curvA{c} = d\G{a'}{A}$
(recall, here, that $\etens{a'}{bc}$ acts as an invariant metric
on the \YM/ Lie algebra $\vsA$). 
Therefore, it follows that 
$\G{a'}{K}$ has the geometrical role of 
a generalized nonabelian \CS/ $3$-form,
determining a generalized Chern class $4$-form 
$\etens{a'}{bc} \J{b}\J{c}$
through the relation \eqref{Chernclass}. 
\end{proposition}

\section{ Geometrical formulation }

In Theorem~4, 
the special cases of 
a pure \YM//\CM/ deformation, 
a pure \FT//\CM/ deformation, 
and pure extended \FT//\CM/ deformations,
each lead to interesting nonlinear theories
with noteworthy geometrical features, 
as will now be described. 
For this purpose it is convenient to employ an index-free notation
and work entirely with $\vsA$-, $\vsB$- valued $1$-forms and $2$-forms
(denoted by bold face field variables). 

\subsection{ \YM//\CM/ theory }

From Theorem~2 and Proposition~3,
the nontrivial algebraic structure for this type of deformation is given by 
a semisimple Lie bracket $[\cdot,\cdot]_\vsA$
and a symmetric bilinear map $e(\cdot,\cdot) =e'\otimes g(\cdot,\cdot)$
where $g(\cdot,\cdot)=tr_\vsA(\cdot\otimes\cdot)$ 
is the Cartan-Killing metric of $\vsA$
and $e'$ is any vector in $\vsB$,
with $b(\cdot)=0$ and $[\cdot,\cdot]_\vsB=0$. 
Accordingly, without loss of essential generality, 
we let $\vsB$ be a normed one-dimensional abelian Lie algebra $\U{1}$
spanned by $e'$,
so the field variables then consist of 
a $\vsA$-valued \YM/ $1$-form $\ymA$
and a $\U{1}$-valued $2$-form $B$. 

The Lagrangian is given by the $4$-form
\EQ
L_{\rm YM/CM} = 
tr_\vsA(*\ymcurvA \ymcurvA) + *(H-e\ymG) (H-e\ymG)
\endEQ
where $\ymcurvA=d\ymA+\frac{1}{2}[\ymA,\ymA]_\vsA$, $H=dB$ are 
the \YM/ curvature $2$-form and abelian field strength $3$-form,
respectively, 
and $\ymG = tr_\vsA( \ymA d\ymA +\frac{1}{3}\ymA [\ymA,\ymA]_\vsA )$
is the nonabelian \CS/ $3$-form \cite{Chernref},
while $e$ is a coupling constant equal to the norm of $e'$. 
This Lagrangian is separately invariant 
under the abelian gauge symmetry
\EQ
\vXvar\ymA =0 ,\quad
\vXvar B = d\vX{} ,
\endEQ
for arbitrary $\U{1}$-valued $1$-forms $\vX{}$,
and under the combined \YM//\CM/ gauge symmetry
\EQ
\ymsXvar\ymA = d\ymsX +[\ymA,\ymsX]_\vsA \equiv \D{\ymA}\ymsX ,\quad
\ymsXvar B =  e\, tr_\vsA( \ymA d\ymsX ) ,
\endEQ
for arbitrary $\vsA$-valued $0$-forms $\ymsX$. 
This theory describes a nonlinear deformation of 
nonabelian \YM/ gauge theory and abelian \FT/ gauge theory for $\ymA,B$.
Its dual formulation involves 
a linear $\U{1}$ sigma (\ie/ scalar) field $\dilaton{}$
introduced through the $B$ field equation by 
\EQ
*H -e {*\ymG} = d\dilaton{} . 
\endEQ
The dual Lagrangian for $\dilaton{}$ and $\ymA$ is given by the $4$-form
\EQ
L^{\rm dual}_ {\rm YM/CM} =
tr_\vsA(*\ymcurvA \ymcurvA) + (2e\ymG \pm {*d}\dilaton{}) d\dilaton{} .
\endEQ
This yields the field equations
\EQ
\D{A} {*\ymcurvA} = -2e \ymcurvA d\dilaton{} ,\quad
d{*d}\dilaton{} = \mp e\, tr_\vsA(\ymcurvA \ymcurvA) .
\endEQ
Hence, $\dilaton{}$ couples to $\ymA$ through 
the nonabelian Chern class $4$-form $tr_\vsA(\ymcurvA \ymcurvA)$ 
\cite{Chernref}.

Thus a pure \YM//\CM/ deformation describes 
an exotic nonabelian \YM/ scalar dilaton theory. 

\subsection{ \FT//\CM/ theory }

For this type of deformation, 
the nontrivial algebraic structure given by Theorem~2 and Proposition~3
consists of a nonabelian Lie bracket $[\cdot,\cdot]_\vsB$
and a symmetric linear map $e^S(\cdot)$ that annihilates 
the commutator ideal $D\vsB=[\vsB,\vsB]$, 
with $b(\cdot)=0$ and $[\cdot,\cdot]_\vsA=0$. 
Consequently, for there to exist a nontrivial such map $e^S(\cdot)$,
it is necessary that $D\vsB$ be a proper Lie subalgebra of $\vsB$
and that the cokernel of $e^S(\cdot)$ belong to 
the orthogonal complement of $D\vsB$ in $\vsA$, 
which we denote by $D^\perp\vsB$. 
Note it then follows that $[D^\perp\vsB,D\vsB] \subseteq D\vsB$
and $e^S(D\vsB)=0$ 
while the linear maps $e^S(u')$ for each $u'$ in $D^\perp\vsB$ are, 
by diagonalization, a sum of one-dimensional projection operators in $\vsA$. 
These algebraic relations are satisfied if, 
with little loss of generality, 
we let $D^\perp\vsB=\U{1}$ and take $\vsB$ to be 
the semidirect product $\U{1}\semi\DvsB$ of 
an abelian Lie algebra $\U{1}$ 
and a semisimple Lie algebra $\DvsB=D\vsB$ 
such that $\U{1}$ and $\DvsB$ are orthogonal subspaces in $\vsB$
with respect to a fixed metric $g'(\cdot,\cdot)$
whose restriction to $\DvsB$ is the Cartan-Killing metric, 
where the $\U{1}$ action on $\DvsB$ is given by 
an adjoint representation $\adDB(e')$ 
using a vector $e'$ in $\DvsB$. 
We also then let $\vsA=\U{1}$
and $e^S(\cdot)=g'(e'_\perp,\cdot) \id_\vsA$
where $e'_\perp$ belongs to $\U{1}$ in $\vsB$
and $\id_\vsA$ is the identity map on $\vsA$. 
Correspondingly, 
the field variables are 
a $\U{1}$-valued $1$-form $A$, 
and a $\U{1}$-valued $2$-form $B$
along with a $\DvsB$-valued $2$-form $\ftB$
as associated with $\vsB=\U{1}\semi\DvsB$. 

To proceed, 
let $F=dA$ and $G=AdA$ denote 
the $\U{1}$ \YM/ (\ie/ Maxwell) field strength $2$-form
and $\U{1}$ \CS/ $3$-form, respectively. 
Introduce as nonlinear \FT/ field strengths
a $\U{1}$-valued $1$-form $K$ and $\DvsB$-valued $1$-form $\ftK$,
which include a \CM/ coupling, 
given by 
\EQ
\ftK = *( \D{\ftK} \ftB + \frac{1}{e} K[e',\ftB]_\DvsB ) ,\quad
K = *(dB + \frac{1}{e} g'(e',[\ftK,\ftB]_\DvsB) -eG ) ,
\endEQ
together with the covariant derivative
\EQ
\D{\ftK} = d +[\ftK,\cdot]_\DvsB ,
\endEQ
where $e=g'(e',e')^{1/2}$ is a coupling constant. 
Now the Lagrangian is given by the $4$-form
\EQ
L_{\rm FT/CM} = 
{*F} dA + tr_\DvsB( \ftK d\ftB ) + K (dB -e G)
\endEQ
with $tr_\DvsB(\cdot\otimes\cdot) = g'(\cdot,\cdot)|_\DvsB$ 
denoting the Cartan-Killing metric on $\DvsB$. 
This Lagrangian is separately invariant (to within an exact $4$-form)
under the \FT/ gauge symmetries
\EQs
&&
\vXvar B = d\vX{} ,\quad
\vXvar \ftB = 0,\quad
\vXvar A = \ftvXvar A =0 ,
\\
&&
\ftvXvar B = \frac{1}{e} g'(e',[\ftK,\ftvX]_\DvsB) ,\quad
\ftvXvar \ftB = \D{\ftK} \ftvX +\frac{1}{e} K[e',\ftvX]_\DvsB ,
\endEQs
for arbitrary $\U{1}$-, $\DvsB$- valued $1$-forms $\vX{},\ftvX$, 
and under the combined Maxwell/\CM/ gauge symmetry
\EQ
\sXvar A = d\sX{} ,\quad
\sXvar B = e F \sX{} ,\quad
\sXvar \ftB = 0 , 
\endEQ
for arbitrary $\U{1}$-valued $0$-forms $\sX{}$. 
This theory describes a nonlinear deformation of 
nonabelian \FT/ gauge theory for $B,\ftB$
and Maxwell (\ie/ $\U{1}$ \YM/) gauge theory for $A$. 

The dual formulation of this theory, 
obtained through the $B,\ftB$ field equations, 
involves both a scalar field $\dilaton{}$
and a chiral (nonlinear sigma) field $U$ 
given by 
\EQ
K = d\dilaton{} ,\quad
\adDB(\ftK) 
 = \chiral\inv d\chiral -e_\DvsB' d\dilaton{}
\endEQ
where $\chiral$ is a matrix belonging to the adjoint representation of
the local semisimple Lie group determined by the Lie algebra $\DvsB$,
and $e_\DvsB'$ denotes the matrix $\frac{1}{e} \adDB(e')$. 
The dual Lagrangian is given by the $4$-form
\EQ
L^{\rm dual}_ {\rm FT/CM} 
= {*F} F +2(e G \pm {*d}\dilaton{}) d\dilaton{} 
\pm tr_\vsB( \chiral\inv {*d}\chiral 
(\chiral\inv d\chiral - 2 e_\DvsB' d\dilaton{}) ), 
\endEQ
which yields the field equations
\EQs
&&
d{*F} = -2e F d\dilaton{} ,\quad
d{*d}\dilaton{} = \mp e FF , 
\\
&&
d{*d}\chiral = d\chiral \chiral\inv {*d}\chiral 
+\chiral [\chiral\inv d\chiral, e_\DvsB'] {*d}\dilaton{}
\mp e FF \chiral e_\DvsB' . 
\endEQs
Hence, $\dilaton{}$ and $\chiral$ couple to $A$ through 
the $\U{1}$ Chern class $4$-form $FF$. 

Therefore a pure \FT//\CM/ deformation describes 
an exotic Maxwell scalar/chiral dilaton theory. 

\subsection{ Extended \FT/ theory }

For an extended \FT/ deformation by itself, 
the most natural algebraic structure is given by 
a semisimple Lie algebra $\vsB$
together with its adjoint representation $b(\cdot)=\adB(\cdot)$
on the vector space $\vsA \simeq \vsB$
(\ie/ $\BintoA=\id$ is a vector space isomorphism), 
and an invariant metric $g(\cdot,\cdot)$ on both $\vsB,\vsA$. 
It is worth describing this type of deformation in more detail, 
since it has some novel features of its own. 
The Lagrangian is given in terms of 
an $\vsA$-valued $1$-form $\ymA$ and 
an $\vsB$-valued $2$-form $\ftB$ by 
\EQ
L_{\rm exFT} = 
tr_\vsA({*\ftJ} \ymF) + tr_\vsB( \ftK \ftH )
\endEQ
where $\ymF=d\ymA$, $\ftH=d\ftB$ are abelian field strengths, 
and the nonlinear \FT/ field strength $2$-form ${*\ftJ}$ and $1$-form $\ftK$
are defined by 
\EQ
\ftJ = \D{\ftK} \ymA ,\quad
\ftK = *( \D{\ftK} \ftB + \adB({*\ftJ}) \ymA ) ,
\endEQ
using the covariant derivative
\EQ
\D{\ftK} = d +\adB(\ftK) .
\endEQ
This Lagrangian is invariant (to within an exact $4$-form)
under the extended \FT/ gauge symmetries
\EQs
&&
\ftvXvar\ymA =0 ,\quad
\ftvXvar\ftB = \D{\ftK} \ftvX , 
\\
&&
\ymsXvar\ymA = \D{\ftK}\ymsX ,\quad
\ymsXvar\ftB = \adB({*\ftJ}) \ymsX , 
\endEQs
for arbitrary $\vsA$-valued $0$-forms $\ymsX$,
$\vsB$-valued $1$-forms $\ftvX$. 
The field equations for $\ymA,\ftB$ 
\EQ
\D{\ftK} {*\ftJ} =0 ,\quad
d\ftK +\frac{1}{2}[\ftK,\ftK]_\vsB =0
\endEQ
yield a dual formulation 
\EQ
d{*d}\chiral = d\chiral \chiral\inv {*d}\chiral ,\quad
d{*d}( \chiral\ymA ) =0 
\endEQ
in terms of the chiral field $\chiral$ 
given by a matrix belonging to the adjoint representation of
the local semisimple Lie group associated with the Lie algebra $\vsB$,
where
\EQ
\ftK = \chiral\inv d\chiral ,\quad
\ftJ = \chiral\inv d(\chiral\ymA) .
\endEQ

Hence, extended \FT/ gauge theory is equivalent to decoupled theories of
a chiral field $\chiral$
and an abelian \YM/ field $\stymA=\chiral\ymA$
where $\stymF=d\stymA$ 
is the abelian \YM/ field strength. 
The structure of this theory, however, is incompatible
with a \CM/ coupling, as will now be demonstrated. 
From Proposition~3,
the inclusion of a \CM/ deformation involves 
a symmetric linear map $e^S(\cdot)$ given by 
either of the two algebraic relations in \Eqref{esol}. 
But, here, notice the first relation becomes trivial, 
$e^S(\cdot) = e b^S(\cdot) =0$, 
since $b(\cdot) = \adB(\cdot) = -\adTB(\cdot)=-b\ad(\cdot)$, 
while the second relation 
$e^S(\cdot) = g(e',\cdot)\id_\vsA$ 
cannot be satisfied because 
there is no nonzero vector $e'$ orthogonal to $D\vsB=[\vsB,\vsB]=\vsB$ 
in a semisimple Lie algebra. 

This proves a no-go result for combining 
a \CM/ deformation with an extended \FT/ deformation
in the natural case where the algebraic structure is based on 
a semisimple Lie algebra $\vsB$. 
However, if a non-semisimple Lie algebra $\vsB$ is considered, 
then there exists a consistent nontrivial extended \FT//\CM/ deformation. 
In particular, the algebraic structure required by Proposition~3
is satisfied by the non-semisimple example for $\vsB$ 
used above in the pure \FT//\CM/ deformation 
if we take 
$b(\cdot) e = e^S(\cdot) =g(e'_\perp,\cdot) \id_\vsA$
for any vector $e'_\perp$ in $\U{1}$, 
where $\vsB$ is the semidirect product $\U{1}\semi\DvsB$
and $\vsA=\U{1}$. 
Here, as above, $\U{1}$ and $\DvsB$ are orthogonal subspaces in $\vsB$,
with the $\U{1}$ action on $\DvsB$ given by 
an adjoint representation $\adDB(e') \equiv e\, e_\DvsB'$ 
involving a fixed vector $e'$ in $\DvsB$,
and a coupling constant $e$ equal to the norm of $e'$. 

This non-semisimple algebraic structure 
provides a generalization of the pure \FT//\CM/ theory for $A,B,\ftB$
to a theory with an extended \FT/ coupling. 
The dual formulation for this deformation is given by just
replacing $F=dA$ with $J=\D{K}A =dA +d\dilaton{} A$
in the Lagrangian $L^{\rm dual}_ {\rm FT/CM}$. 
Under this replacement, 
since $\vsA=\U{1}$ is abelian, 
the $\U{1}$-valued $3$-form 
\EQ
\frac{1}{e} \G{}{K}=A\D{K}A =A( dA +d\dilaton{} A ) =AdA =\G{}{A} 
\endEQ
is equal to the $\U{1}$ \CS/ $3$-form $\G{}{A}$. 
The resulting field equations are given by 
\EQs
&&
d{*J} = ({*J} -2e J)d\dilaton{} ,\quad 
\pm d{*d}\dilaton{} = {*J}J -e JJ ,
\\
&&
d{*d}\chiral = 
d\chiral \chiral\inv {*d}\chiral 
\pm ( {*J}J -eJJ )\chiral e_\DvsB' . 
\endEQs
Hence, in this theory, 
the abelian $1$-form $A$ couples to a scalar dilaton $\dilaton{}$ 
and a chiral dilaton $\chiral$ 
through a generalized $\U{1}$ Chern class $4$-form 
$JJ =FF-2\G{}{A}d\dilaton{}$
as well as a standard Maxwell type $4$-form ${*J}J$,
which are related by $d(e\G{}{A}-A{*\D{K}A}) =eJJ-{*J}J$
on solutions $A$. 
In particular, since $\D{K} =d+d\dilaton{}$, it follows that 
$J=e^{-\dilaton{}} \stJ{}$ and $A=e^{-\dilaton{}} \stA{}$
where $\stJ{}=d\stA{}$ is the Maxwell field strength of $\stA{}$.
Thus, the effect of adding an extended \FT/ coupling 
produces a standard dilaton coupling of $\dilaton{}$ to $A$
in the pure \FT//\CM/ deformation. 

Finally, the geometrical structure of this theory
illustrates a general feature that 
the \CM/ and (extended) \FT/ deformations possess opposite parity. 
Specifically, 
under a parity operator defined by
$\parity=\parity{}^2$ such that 
$\parity d=d\parity$ 
and $\parity *=-{*\parity}$,
it follows from Proposition~1 that 
if $\A{a} \rightarrow \parity\A{a}$, $\B{a'} \rightarrow \parity\B{a'}$, 
then $\nthL{3}{}_{\rm YM} \rightarrow -\parity\nthL{3}{}_{\rm YM}$, 
$\nthL{3}{}_{\rm FT} + \nthL{3}{}_{\rm exFT} 
\rightarrow \parity(\nthL{3}{}_{\rm FT} + \nthL{3}{}_{\rm exFT})$,
and $\nthL{3}{}_{\rm CM} \rightarrow -\parity\nthL{3}{}_{\rm CM}$.

\section{ Concluding remarks }

The most general class of covariant nonabelian gauge theories of 
a set of coupled massless $1$-form and $2$-form fields in four dimensions
has been constructed in this Letter, 
arising from the recent classification of 
geometrical nonlinear deformations of 
the linear abelian gauge theory of such fields
\cite{HenneauxKnaepen1,HenneauxKnaepen2,YMFTpaper,geometrical}.
These deformations comprise 
a Yang-Mills coupling of the $1$-forms,
a Freedman-Townsend coupling of the $2$-forms,
an extended Freedman-Townsend type coupling 
between the $1$-forms and $2$-forms,
in addition to a Chapline-Manton type coupling of 
the $1$-forms with the $2$-forms. 
The well known duality \cite{FTth} of 
nonabelian \FT/ gauge field theory of $2$-forms
and nonlinear sigma field theory for Lie group targets
carries over to the more general gauge field theories 
of $1$-forms and $2$-forms
presented here. 
In particular, due to the presence of the Chapline-Manton coupling,
the dual formulation of these theories is found to describe
a class of exotic \YM/ dilaton theories
in which the nonlinear sigma field has a novel type of dilaton coupling
to the \YM/ $1$-form field
through a generalized Chern class term. 
This \YM/ dilaton coupling gives an interesting generalization of
nonlinear sigma field theory into Lie group targets
on four dimensional manifolds. 

In the case of a Euclidean manifold (namely, $*^2=\openone$),
the \YM/ dilaton field equations represent harmonic maps 
coupled to elliptic equations for a \YM/ connection 
(modulo gauge conditions), 
while in the case of a Lorentzian manifold (namely, $*^2=-\openone$),
the field equations instead represent wave maps 
coupled to hyperbolic equations for a \YM/ connection 
(modulo gauge conditions). 
Since the field equations in both cases 
involve a generalized Chern class term,
the solutions of the resulting elliptic and hyperbolic nonlinear systems
may be expected to possess some analytic features that depend on 
global properties of the bundle of \YM/ connections
and the underlying four-manifold. 
As such, these systems should prove to be of interest of study 
in areas of mathematical physics related to 
the \YM/ equations, harmonic maps and wave maps.


\begin{thebibliography}{99}
\bibitem{Chernclass}
In pure \YM/ theory on a $n$-dimensional manifold $M$
with a gauge group $\group$, 
the Chern class is the gauge invariant $n$-form 
given by the exterior product of
the \YM/ curvature $2$-form and its $n-2$-form Hodge dual,
contracted with the Cartan-Killing metric of 
the Lie algebra $\vsA$ of $\group$. 
The integral of the Chern class over $M$ 
defines a global differential invariant,
associated with the principle bundle on $M$ 
whose structure group is $\group$.
See, \eg/, \Ref{Chernref}.
\bibitem{geometrical}
Here, geometrical deformations mean that 
all deformation terms are locally constructed in a covariant manner
entirely from the $p$-form fields 
and their associated gauge symmetry parameters 
(arbitrary $p-1$-forms on $M$),
using only the Hodge dual $*$ and exterior derivative $d$ operators.
\end{thebibliography}
\end{document}